\documentclass[aps,prd, twoside,superscriptaddress,showpacs, preprintnumbers, showpacs, nofootinbib, 
twocolumn]{revtex4-1}
\pdfoutput=1

\usepackage{amssymb,amsmath,latexsym,graphics, graphicx,epsfig,multirow,comment,hyperref,appendix, verbatim} 
\usepackage{pifont}
\usepackage{lipsum}
\usepackage{mathrsfs}  

\usepackage{slashed}
\usepackage{subfigure}
\usepackage{feynmp}
\usepackage{graphicx}
\usepackage{amsfonts}
\usepackage{color}
\usepackage{cancel}

\def\hbar{\hspace{0pt}\raisebox{1pt}{$-$} \hspace{-7pt} h}

\def\5{\overline 5}

\definecolor{JJ}{RGB}{0,144,255}

\newcommand{\be}{\begin{equation}}
\newcommand{\ee}{\end{equation}}
\newcommand{\bea}{\begin{eqnarray}}
\newcommand{\eea}{\end{eqnarray}}

\newcommand{\ba}{\begin{eqnarray}}
\newcommand{\ea}{\end{eqnarray}}

\begin{document}
\title{On the Direct Detection of Dark Matter Annihilation}

\author{John F. Cherry}

\affiliation{Theoretical Division, Los Alamos National Laboratory, Los Alamos, New Mexico 87545, USA}
\author{Mads T. Frandsen}

\author{Ian M. Shoemaker}
\affiliation{CP$^{3}$-Origins and the Danish Institute for Advanced Study, University of Southern Denmark, Campusvej 55, DK-5230 Odense M, Denmark}


\begin{abstract}

We investigate the direct detection phenomenology of a class of dark matter (DM) models in which DM does not directly interact with nuclei, {but rather} the products of its annihilation do. When these annihilation products are very light compared to the DM mass, the scattering in direct detection experiments is controlled by relativistic kinematics. 
This results in a distinctive recoil spectrum, a non-standard and or even {\it absent} annual modulation, and the ability to probe DM masses as low as a $\sim$10 MeV. We use current LUX data to show that experimental sensitivity to thermal relic annihilation cross sections has already been reached in a class of models.  Moreover, the compatibility of dark matter direct detection experiments can be compared directly in $E_{{\rm min}}$ space without making assumptions about {DM astrophysics, mass, or scattering form factors}. Lastly, when DM has direct couplings to nuclei, the limit from annihilation to relativistic particles in the Sun can be stronger than that of conventional non-relativistic direct detection by more than three orders of magnitude for masses in a 2-7 GeV window.

\preprint{CP3-Origins-2015-002 DNRF90, DIAS-2015-2, LA-UR-15-20058.}

\end{abstract}




\maketitle





 \textbf{\textit{Introduction -}} 
While very little is known about Dark Matter (DM), its cosmological abundance is experimentally quite well-determined: $\Omega_{CDM}h^{2} = 0.1199 \pm 0.0027$~\cite{Ade:2013zuv}.   {An appealing} 
framework for understanding the relic abundance of Dark Matter (DM) is thermal freeze-out~\cite{zeldo}. Number-changing interactions in the early universe, $\overline{X}X \leftrightarrow (\overline{{\rm SM}}) {\rm SM}$ keep DM in thermal equilibrium with the SM bath, until the rate of these annihilation processes drops below the rate of Hubble expansion. After this point the abundance of DM is essentially fixed at, $\Omega_{CDM}h^{2}  \simeq 0.12  \left(6 \times 10^{-26}~{\rm cm}^{3}~{\rm s}^{-1}/\langle \sigma_{ann}v_{rel}\rangle\right)$, singling out a characteristic annihilation cross section $\langle \sigma_{ann}v_{rel}\rangle$ for thermally produced DM to yield the observed abundance.  This scenario is attractive in that it provides a simple and elegant framework for the relic abundance that can be tested in a variety of ways, including direct detection (DD)~\cite{Goodman:1984dc}.  {However, current constraints from DD rule out many of the simplest models of thermal relic DM, which may indicate a modification of the above picture.} 

In this paper we investigate a modification of thermal DM which alleviates the tension between DD constraints and the thermal relic hypothesis, while making unique predictions for DD.  In particular, we take the abundance of DM, $X$, to be determined by the annihilation process $\overline{X}X \leftrightarrow \overline{Y}Y$, where $Y$ is a much lighter dark sector species.  The interactions of the dark sector state $Y$ with ordinary nuclei allows for a unique test of the scenario at DD experiments.  
The resulting DD phenomenology of this class of models is distinctive, owing to the fact that (1) the scattering partner of the nucleus is relativistic, rendering the kinematics of scattering completely different and (2) it is the flux of the scattering partner $Y$ that determines the rate of events at a detector rather than $X$. Both of these features have novel consequences not considered in the literature of ``model- independent'' direct detection analyses~\cite{Fitzpatrick:2012ix,DelNobile:2013sia}.  

As loop processes will always engender scattering of $X$ on nuclei at DD, we will focus on DM masses less than $\sim$GeV such that the non-relativistic scattering of $X$ does not produce detectable nuclear recoils above a detector's $\mathcal{O}({\rm keV})$ threshold. Similar scenarios have recently been investigated in~\cite{Huang:2013xfa,Agashe:2014yua,Berger:2014sqa,Kong:2014mia} with a focus on the Cherenkov signals at Super-Kamiokande and IceCube.

In this paper we employ current LUX~\cite{Akerib:2013tjd} limits to demonstrate that DD experiments are sensitive to thermal relic annihilation cross sections for galactic center annihilation of DM in a window of DM masses from 10 MeV to 1 GeV.  Direct detection has historically been muddled by multiple conflicting data sets.  To combat this, we illustrate how current and future direct detection data can be easily analyzed for compatibility in this framework by mapping results to $E_{{\rm min}}$-space.  Additionally, we investigate the testability of such relativistic scattering models where the signal is dominated by DM accretion and annihilation within the sun.

\textbf{\textit{Annihilating DM in the Galactic Center - }} 
Two potential sources of DM annihilation 
are annihilation from the Galactic Center and annihilation within the Sun.  A key difference between these two is that the {latter} relies on {a stable balance between the accretion and evaporation rates of DM interacting with nucleons inside the Sun}. We first consider galactic center annihilation since this does not require a build up of DM in the Sun and hence requires fewer assumptions.   

For simplicity consider ``2-to-2'' annihilation, $\overline{X}X \rightarrow \overline{Y}Y$. 
Then the differential rate (per unit detector mass) at a direct detection experiment is,
\be 
\frac{dR}{dE_{R}} = \frac{\Phi_{Y}}{m_{N}} \int_{E_{min}(E_{R})}^{\infty} dE_{Y} \frac{d N}{dE_{Y}} \left(\frac{d\sigma_{YN}}{d E_{R}}\right)\, ,
\label{eq:rate}
\ee
\\
where $\Phi_{Y}$ is the local flux of $Y$'s, $E_{min}(E_{R}) = \sqrt{m_{N}E_{R}/2}$ is the minimum energy to produce a recoil of energy $E_{R}$, and $\frac{dN}{dE_{Y}} = 2 \delta \left(E_{Y} - m_{X}\right)$. 

\begin{figure}[t!] 
\begin{center}
 \includegraphics[width=.4\textwidth]{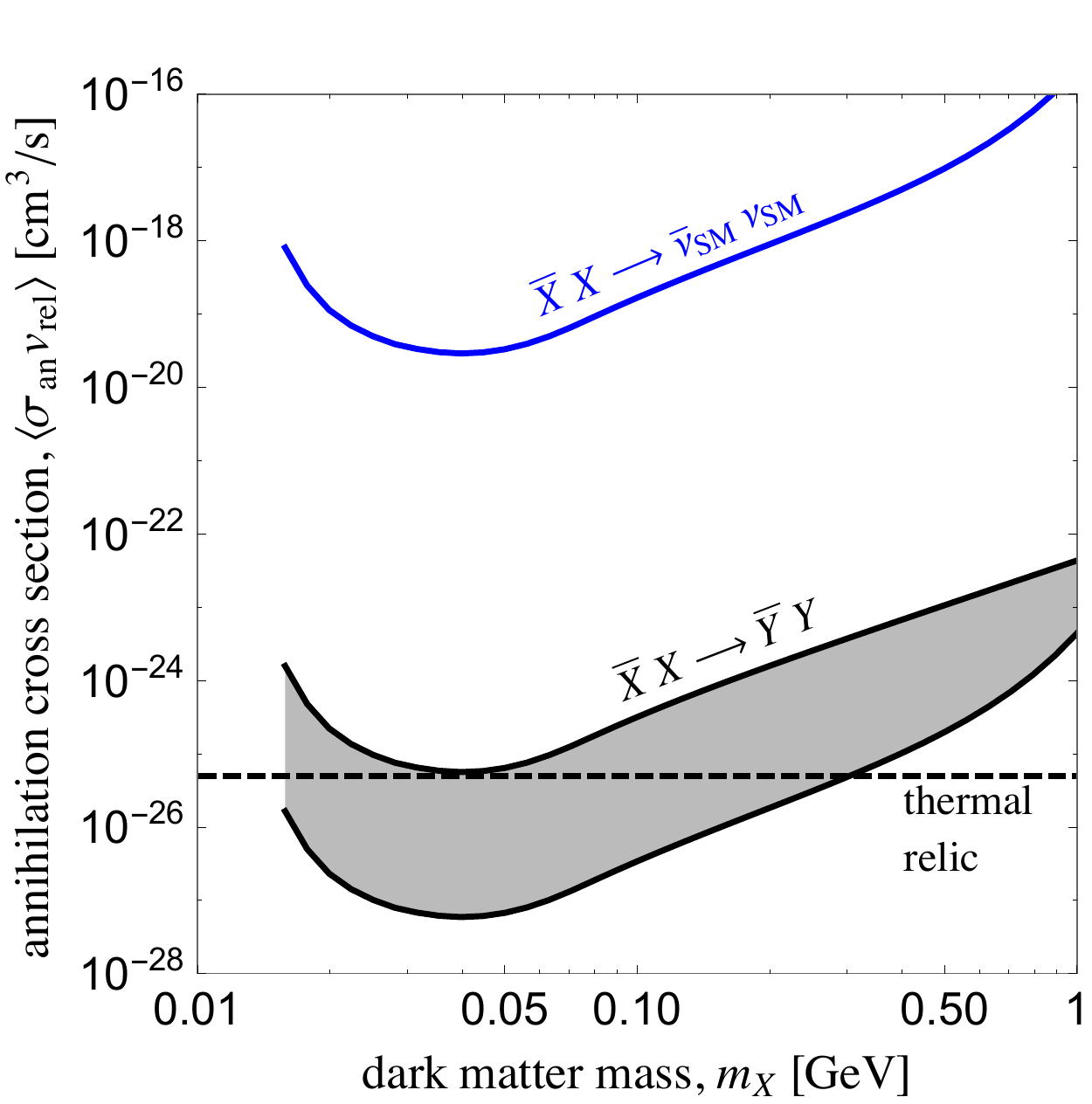}
\caption{{\it }: {LUX Limits on $\bar{X} X \rightarrow \overline{\nu}_{{\rm SM}} \nu_{{\rm SM}}$, for which $G_{Y} = G_{F} = 1.2\times10^{-5}~{\rm GeV}^{-2}$, and a model in which DM annihilates to relativistic pairs  $~\overline{Y}Y$ which scatter on nuclei, $YN \rightarrow YN$, via Eq.(\ref{eq:model}). In the shaded gray band we vary $G_{Y}$ in the interval $\left(7\times 10^{3}-5\times 10^{2}\right)G_{F}$.} Propagation effects have been accounted for in determining the flux of $Y$ at the underground site of the LUX detector (see Appendix).}
\label{fig:nuDM}
\end{center}
\end{figure}

For simplicity, {we adopt} a contact interaction between $Y$ and a quark $q$ of the form $\mathcal{O}_{qY} = G_{Y} \left(\bar{Y} \gamma_{\mu} Y\right) \left(\bar{q}\gamma^{\mu}q\right)$ where $G_{Y}$ is the effective coupling. By analogy with neutrino-nucleus elastic~\footnote{Note that in~\cite{Pospelov:2011ha} it was estimated that the inelastic scattering cross section is small at low-energies compared to the elastic cross section, $\sigma({\rm elastic}) /\sigma({\rm inelastic})\sim A^{2}/(E_{Y}^{4}R_{N}^{4})$, where  {the nuclear radius} is, $R_{N} \sim (10~{\rm MeV})^{-1}$. Thus for a Xenon target nucleus, inelastic scattering is sub-dominant for $E_{Y} \lesssim$ GeV. Given our focus on sub-GeV DM we will ignore inelastic processes in this paper.} scattering
the differential cross section is~\cite{Pospelov:2011ha},
\be 
\frac{d \sigma_{YN}}{dE_{R}} = \frac{G_{Y}^{2}}{2\pi} A^{2} m_{N} F^{2}(E_{R})~\left[ 1 - \left(\frac{E_{min}}{E_{Y}}\right)^{2} \right]\, ,
\label{eq:diffcx}
\ee
{where $F(E_{R})$ is Helm's nuclear form factor~\cite{Helm:1956zz}, and the $A^{2}$ coefficient is for the coherent enhancement of scattering with equal rates on protons and neutrons. }

{Next we must determine the local flux of $Y$}. The flux of $Y$ particles from DM annihilation {in the Galactic Center} is estimated as~\cite{Agashe:2014yua}, $\Phi_{Y}  =   1.6 \times 10^{-2} {\rm cm}^{-2}{\rm s}^{-1} \left(\frac{ \langle \sigma_{\bar{X}X\rightarrow \bar{Y}Y}v_{rel}\rangle}{5 \times 10^{-26}~{\rm cm}^{3}{\rm s}^{-1}}\right) \left(\frac{20~{\rm MeV}}{m_{X}}\right)^{2}$. 

Given this {flux}, and a model of $Y$-nucleus interactions, the only remaining parameter to determine is the annihilation cross section, {which we take to be a free parameter, to be determined from data}. First let us take a minimal choice {by relying} on the SM to furnish the interactions of $Y$ with the nucleus. This immediately singles out the neutrinos as the only {SM} possibility for $Y$. The elastic, spin-independent scattering of SM neutrinos with nuclei can be computed using Eq.~\ref{eq:diffcx} with the replacement, $G_{Y}A^2 \longrightarrow G_{F}(N/2)^{2}$, where $N$ is the number of neutrons and $G_{F}$ is the Fermi constant.  We see {in Fig.~\ref{fig:nuDM}} that with present LUX data, the resulting sensitivity to the annihilation cross section is weak, being orders of magnitude away from thermal relic sensitivity.

{On the other hand, DM could well annihilate to non-SM particles that have larger than electroweak-size interactions.   {Two generic classes of models serve as examples: models of gauged baryon number~\cite{Rajpoot:1989jb,He:1989mi,Foot:1989ts,Carone:1994aa,Aranda:1998fr,FileviezPerez:2010gw,Dulaney:2010dj,Graesser:2011vj,Duerr:2013dza,Tulin:2014tya} and so-called ``Higgs portal'' models~\cite{McDonald:1993ex,Burgess:2000yq,Patt:2006fw,Andreas:2008xy,Andreas:2010dz,Djouadi:2011aa,Pospelov:2011yp,Greljo:2013wja,Bhattacherjee:2013jca,Cline:2013gha}.  Gauged baryon number is motivated by the stability of the proton, which in the SM remains a mystery and may indicate that baryon number is in fact a gauge symmetry.  This is one of the few phenomenologically viable ``portals'' connecting the dark and visible sectors, as it does violate any of the approximate symmetries of the SM. In addition, the Higgs portal, $\mathscr{L} \supset |\phi|^{2} |H|^{2}$ (where $H$ is the SM Higgs and $\phi$ a dark sector scalar), represents a rather generic possibility for connecting the dark and visible sectors. In this case it is natural for scattering on nuclei to be enhanced relative to electrons since the couplings scale with the SM-Higgs Yukawa couplings. }

For illustration, we can make use of a simplified model for quark-$Y$ interactions via the exchange of a light vector $V_{\mu}$ or scalar $\phi$:
\bea 
\mathscr{L}_{V} &\supset& \phi^{\mu} \left(g_{q} \overline{q}\gamma_{\mu} q + g_{Y}\overline{Y} \gamma_{\mu} Y\right),\\
\mathscr{L}_{S} &\supset& \phi \left( g_{q} \overline{q}q +g_{Y} \overline{Y}  Y\right),
\label{eq:model}
\eea
where $g_{Y},g_{q}$ are the couplings of SM quarks and $Y$ to the mediator. In terms of these couplings, the effective parameter is, $G_{Y}= \left(g_{q}g_{Y}\right)/m_{\phi}^{2}$ and the relevant constraints for the two models are discussed respectively in the Appendix.

{As a benchmark we take $G_{Y}$ in Fig.~\ref{fig:nuDM} to vary in the interval $\left(7\times 10^{3}-5\times 10^{2}\right)G_{F}$.  We highlight that the values of $G_{Y}$ are well within the constraints allowed by ``missing energy'' collider limits~\cite{Shoemaker:2011vi,Giardino:2013bma}. Larger $G_{Y}$ are permitted by collider limits, though the flux of $Y$ particles becomes strongly suppressed (see the Appendix for a discussion of this effect).}}  
Nonetheless, we see in Fig.~\ref{fig:nuDM} that models of this type are already being probed by direct detection and can in particular exclude thermal relics in the 10 MeV - 0.5 GeV window, yielding a novel probe of thermal DM. Future constraints will cut further into thermal relic territory.

It is important to observe that annihilation of DM to relativistic states from the galactic center predicts no sizeable annual modulation. 
In the case of annihilation from the Sun however, the annual modulation is known to peak in January due to the eccentricity of the Earth's orbit. Thus solar neutrino signals in direct detection experiments predict a nearly maximally ``wrong'' phase\footnote{ {We note that the phase can be reversed for heavy DM at a high threshold experiment.}} with respect to the expectation from  {light} non-relativistic DM of June 2$^{{\rm nd}}$~\cite{Davis:2014cja}. This expectation can be violated however when the annihilation product $Y$ experiences flavor oscillations on $\mathcal{O}({\rm AU})$ length scales as in for example~\cite{Pospelov:2011ha,Pospelov:2012gm,Pospelov:2013rha} though this requires very small mass-splittings, $\Delta m^{2} \sim 10^{-10}~{\rm eV}^{2}$.

{
It is important to observe that the ability to scatter on nuclei does not induce any physics which would allow $Y$ to decay. Given the generic stability of both $X$ and $Y$, we must be sure that their total abundance does not exceed the observed value, $\Omega_{CDM}h^{2}\simeq 0.2$. For simplicity, we will work in the limit that $Y$ forms a sub-dominant component, i.e. $\Omega_{Y} \ll \Omega_{X} \simeq \Omega_{CDM}$.  This can be naturally arranged when $Y$ is similar to a neutrino and freezes-out when it is relativistic, i.e. $m_{Y} \lesssim {\rm eV}$. Non-relativistic freeze-out of $Y$ can also lead to a small relic abundance when its annihilation cross section is large~\cite{Agashe:2014yua}.    }

{Another potential constraint on these models is the additional radiation energy density they generate during the Big Bang, parameterized by, $\Delta N_{{\rm eff}} = \frac{\rho_{Y}}{\rho_{\nu}} = \frac{g_{Y}T^4_{Y}}{g_\nu T^4_\nu}$, where the photon and neutrino temperatures are related by, $T_\nu = \left( 4/11\right)^{1/3} T_\gamma$.  Due to the annihilation of degrees of freedom from the standard model plasma, the temperature of the dark sector relative to the standard model sector is suppressed via dilution, $T_{dark} = \left( g_{\star,\ \rm post-BBN}/g_{\star,\ \rm dark\ freeze\ out}\right)^{1/3}T_{\gamma}$, where the number of standard model degrees of freedom (DOFs) after Big Bang Nucleosynthesis (BBN) is $g_{\star,\ \rm post-BBN} = 3.36$, the number of standard model DOFs after dark freeze out is $g_{\star,\ \rm dark\ freeze\ out}$, and $T_\gamma$ is the photon temperature.  Assuming $Y$ is the lightest stable particle in the dark sector, so that heavier dark sector DOFs re-heat the bath of $Y$ radiation as they annihilate.  Under the assumption that entropy is conserved, $ g_{DS} T^3_{dark} = g_{Y}T^3_{Y} $, where $g_{DS}$ count the total DOFs in the dark sector. 
Typical values of $\Delta N_{{\rm eff}} \simeq  0.2$ for $g_{\star,\ \rm dark\ freeze\ out}$ corresponding to the QCD phase transition and minimal additional degrees of freedom in the dark sector (only $X$ and $Y$). This is well within the allowed constraints on $\Delta N_{{\rm eff}}$~\cite{Ade:2013zuv,Steigman:2012ve}.}

\textbf{\textit{Direct Detection in $E_{\rm min}$-space - }} 
Direct detection involves a unique combination of particle physics, nuclear physics, and astrophysics. The kinematics of scattering in the non-relativistic case are controlled by the minimum {DM particle} velocity,  $v_{{\rm min}} (E_{R})$, required to produce a nuclear recoil of energy $E_{R}$.  In the absence of unknown form factors, all experimental data can be mapped into $v_{{\rm min}}$-space at each DM mass and compared without specifying the nature of the astrophysical distribution or density of DM~\cite{Fox:2010bu,Fox:2010bz}. These ``halo-independent'' methods have received significant attention~\cite{Frandsen:2011gi,Gondolo:2012rs,Frandsen:2013cna,DelNobile:2013cta,Bozorgnia:2013hsa,DelNobile:2013cva,DelNobile:2013gba,Fox:2013pia,DelNobile:2014eta,Feldstein:2014gza,Fox:2014kua,Gelmini:2014psa,Scopel:2014kba,Cherry:2014wia,Feldstein:2014ufa,Bozorgnia:2014gsa}.   We generalize these methods to cover relativistic scattering as well, where the ``halo-independence'' here comes from the absence of specific assumptions regarding the local DM density, density profile, velocity distribution, and annihilation source.

\begin{figure}[t!] 
\begin{center}
 \includegraphics[width=.4\textwidth]{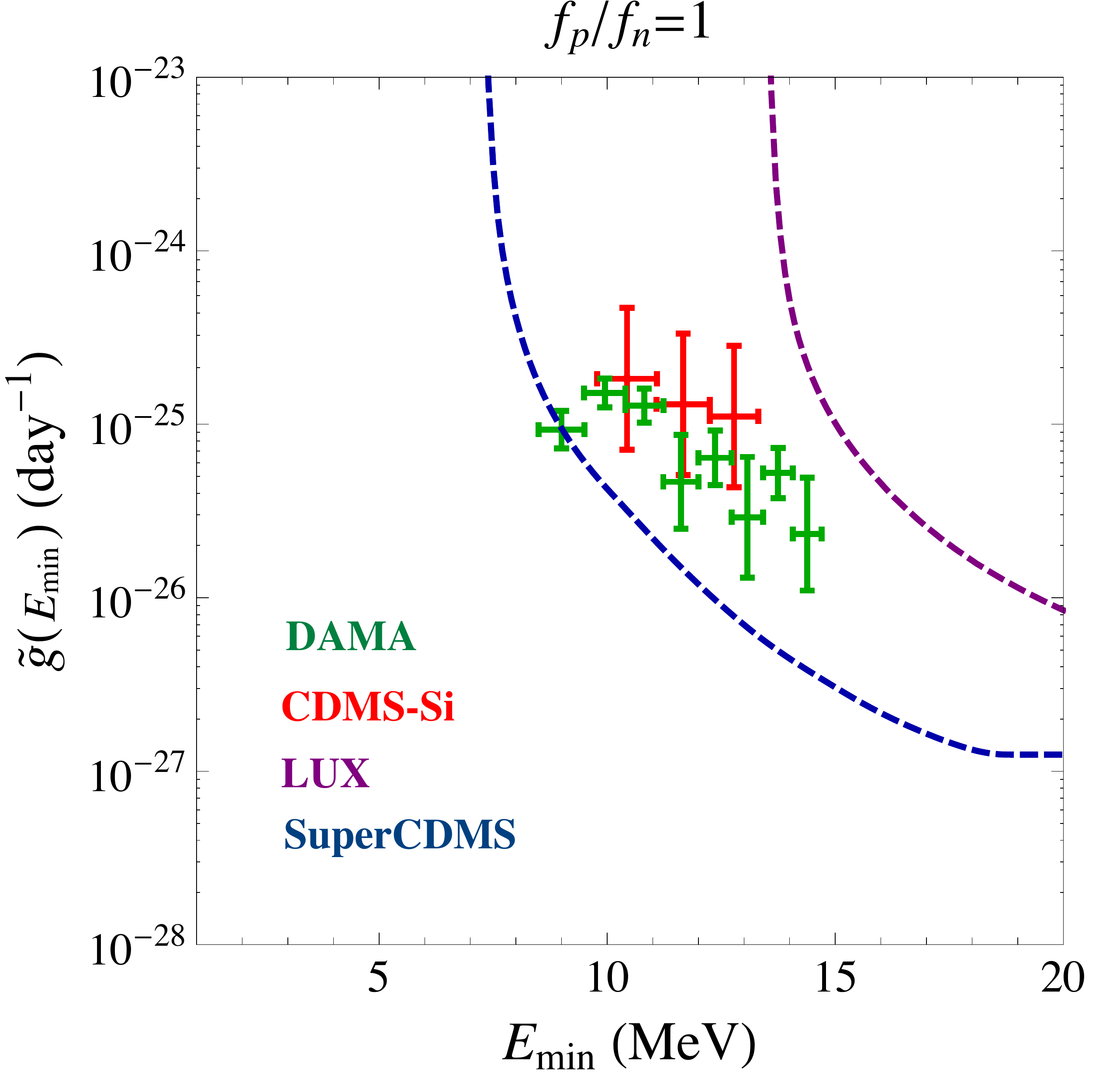}
\caption{{\it } DM mass independent comparison of direct detection data under the assumption of relativistic scattering. Here we include DAMA modulation amplitude from~\cite{Bernabei:2010mq} and the 3 CDMS-Si events~\cite{Agnese:2013rvf}, along with constraints from LUX~\cite{Akerib:2013tjd} and SuperCDMS~\cite{Agnese:2014aze}. }
\label{fig:discrepancy}
\end{center}
\end{figure}

As can be seen from Eq.~(\ref{eq:rate}) and the form of $E_{min}(E_{R})$, the relativistic scattering case allows a comparison of data which is {\it independent of the DM mass}.  {In the case of a claimed detection, using Eq.~(\ref{eq:rate}) we can divide out the nucleus-specific quantities
\be
\tilde{g}(E_{\rm min}) \equiv 2\mu_{n}^{2} (A^{2} F^{2}(E_{R}))^{-1} dR/dE_{R},
\ee
to immediately obtain the preferred $\tilde{g}$ range in Fig.~\ref{fig:discrepancy}. Finally, since the integrand in Eq.~\ref{eq:rate} is strictly positive we can derive conservative limits on $\tilde{g}$ as in~\cite{Fox:2010bz} by assuming a step function form for $\tilde{g}(E_{{\rm min}})$.}   {One can view this procedure as mapping direct detection rates to the $\left(\tilde{g}-E_{{\rm min}}\right)$ plane, which we refer to as ``$E_{{\rm min}}$-space'' for brevity.}  {T}he form of $E_{{\rm min}}(E_{R})$ has the interesting effect of strongly suppressing the sensitivity of experiments employing heavy target nuclei. It is also interesting to observe that LUX~\cite{Akerib:2013tjd} and a relativistic DM interpretation of DAMA~\cite{Bernabei:2010mq} and CDMS-Si~\cite{Agnese:2013rvf} data are fully compatible, though essentially ruled out by the recent SuperCDMS data~\cite{Agnese:2014aze}. 
Clearly, allowing for isospin-violation in order to suppress the sensitivity from Germanium-scattering would result in the positive signals seen by CDMS-Si and DAMA and the null results of LUX and SuperCDMS to be compatible. 

 {Let us pause to highlight the relevance and generality of the halo-independent method employed here. In contrast with non-relativistic scattering, here the velocity distribution matters very little for the rate of events. However, now the astrophysical uncertainty is more fundamental in the sense that the {\it source} of the flux is unknown, i.e. the Galactic Center, the solar interior, etc. Moreover, even after specifying a source there exist large uncertainties in the spatial distribution. This method is independent of these sizeable uncertainties.  Finally, in addition to being DM mass independent, this method is also automatically independent of the energy dependence of the $Y$-nucleus scattering and the spectrum of the $Y$ particles. The generality of this method is illustrated by the model of ``baryonic neutrinos'' which was proposed to account for DAMA's annual modulation~\cite{Pospelov:2011ha,Harnik:2012ni,Pospelov:2012gm,Pospelov:2013rha}. In this case, despite the fact that the source of the relativistic $Y$'s is completely disconnected from DM, the $E_{{\rm min}}$-space representation is valid and allows for a complete comparison of experiments as in Fig.~\ref{fig:discrepancy}.}

\begin{figure}[b!] 
\begin{center}
 \includegraphics[width=.4\textwidth]{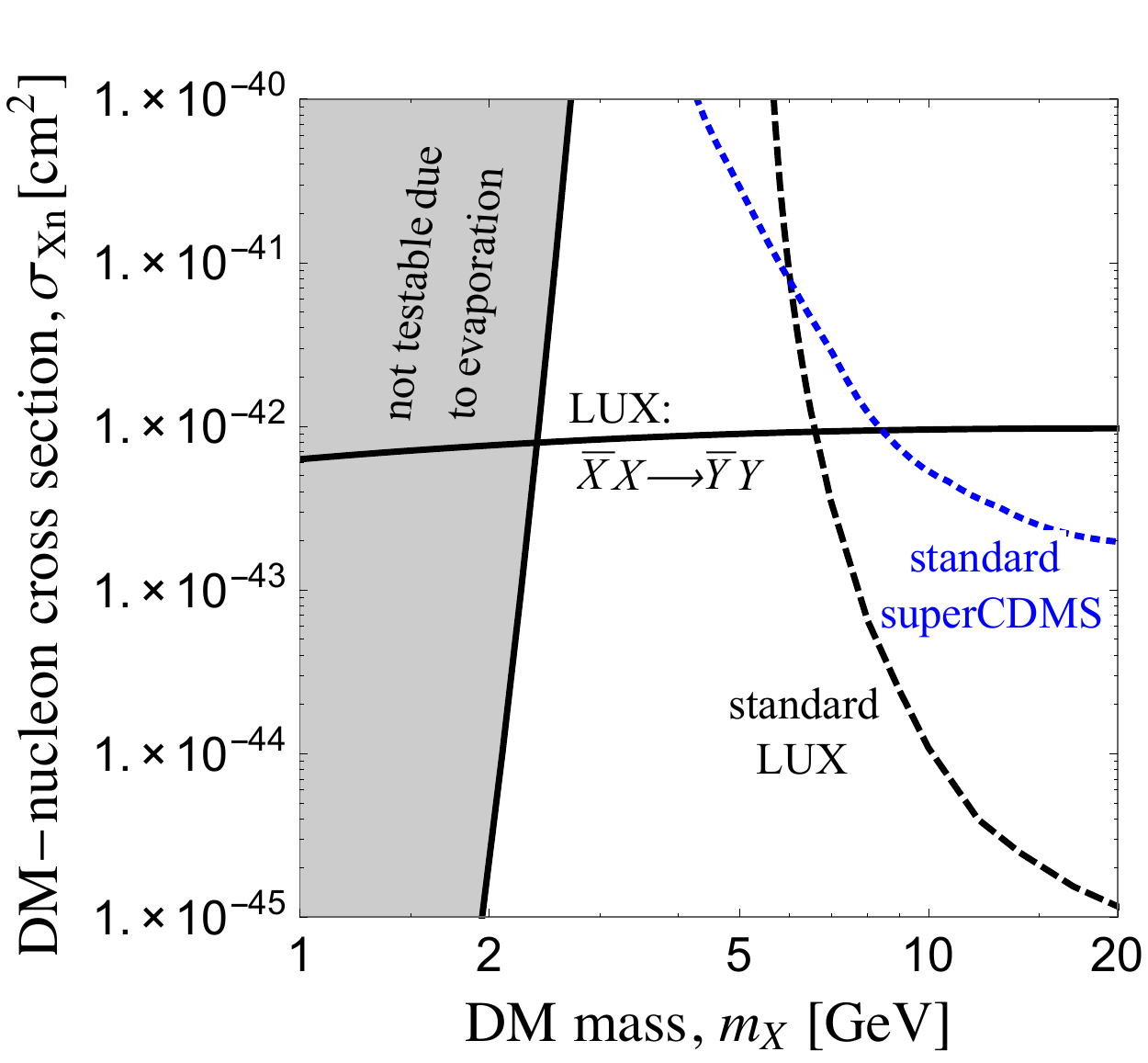}
\caption{{\it } Probing DM-proton interactions from the annihilation in the Sun. For reference we show the cross section limits from LUX and SuperCDMS derived under the assumption of ordinary non-realistic scattering. Here the scattering of the relativistic state is via vector with $G_{Y}=7\times 10^{3}G_{F}$.}
\label{fig:heavyDM}
\end{center}
\end{figure}

\textbf{\textit{DM Annihilation from the Sun - }} 
 {For solar annihilation to dominate over the contribution from the {Galactic Center}, the Sun must contain a large quantity of captured DM. In a symmetric DM context, the solar DM abundance has the time evolution $\dot{N}_{X} = C_{XN} - C_{A} N_{X}^{2}$, such that solar annihilation fluxes are roughly $\Phi_{\odot} \simeq \frac{C_{A}N_{X}^{2}}{4\pi R_{AU}^{2}}$, where $C_{A}$ is the annihilation rate, $N_{X}$ is the number of DM particles in the Sun, and $R_{AU}$ is the Earth-Sun distance.} Assuming that only annihilation and nuclear capture play a role we can specify a model of scattering of the $Y$ states on nuclei and then derive bounds on the DM-nucleus cross section.  {This assumption is valid in the regime where evaporation {of DM out of the sun} is negligible, $N_{eq} \simeq \sqrt{C_{XN}/C_{A}}$, such that the flux depends only on the capture rate. }

 {We again adopt the model of DM annihilating to dark sector $Y$ states that interact with quarks via Eq. (\ref{eq:model}) and take $g_{Y}=0.1$ and  $m_{V} = 50$ MeV.} This yields the result shown in Fig.~\ref{fig:heavyDM}. At low DM mass the limit cannot be trusted, {as} sufficiently light DM is prone to evaporation from collisions with solar nuclei~\cite{Gould:1987ir,Gould:1989tu,Busoni:2013kaa}. 
We note that in models where DM experiences significant self-interaction the abundance of DM in the Sun can be much larger, which can strengthen the limit in Fig.~\ref{fig:heavyDM} significantly~\cite{Zentner:2009is,Frandsen:2010yj,Fan:2013bea,Chen:2014oaa}. We also leave for future work the extension of the framework considered here to an asymmetric DM scenario (see e.g.~\cite{Graesser:2011wi,Lin:2011gj,Bell:2014xta} and \cite{Fan:2013bea}).

\textbf{\textit{Discussion and Summary- }} In summary, this work has investigated the sensitivity of direct detection searches to dark matter annihilation.  Thermal relic dark matter sets a natural scale for the thermally averaged DM annihilation cross section around $\langle \sigma_{ann} v_{rel}\rangle \simeq6\times 10^{-26}~{\rm cm}^{3}~{\rm s}^{-1}$. This scale can be searched for in CMB~\cite{Madhavacheril:2013cna}, gamma-ray~\cite{Ackermann:2013yva}, and even neutrino data~\cite{Aartsen:2013dxa}.  Both CMB and gamma-ray data have breached thermal relic sensitivity for light DM masses.  Though these constraints have sizable astrophysical uncertainties, they may indicate that light DM requires non-SM modes of annihilation.  Here we have studied models in which DM annihilates to a light, non-SM state that can scatter elastically on nuclei and deposit a detectable recoil energy.   {Models of the type considered here retain the appeal of the thermal relic hypothesis while remaining experimentally verifiable}.  We have furthermore demonstrated that in this class of indirect annihilation searches, all astrophysical uncertainties can be ``integrated out''~\cite{Fox:2010bz} and experimental sensitivities can be directly compared. 

This work could be extended to include electronic scattering at direct detection~\cite{Essig:2012yx}, though the reduction in the Cherenkov threshold for electrons implies that Super-K limits extend to much lower masses for leptophilic models~\cite{Agashe:2014yua}. The most similar stud{ies} to our own {which have} been recently carried out assumed that DM interacts with the SM through a kinetically mixed photon, implying both hadronic and electronic couplings~\cite{Agashe:2014yua,Berger:2014sqa}. In this case, large volume detectors like Super-Kamiokande and IceCube  yield very strong limits. {In contrast,} we are interested in a complementary portion of the parameter space compared to~\cite{Agashe:2014yua,Berger:2014sqa} in that we have focused on hadronic models where: (1) the annihilation products are nearly massless compared to nuclear recoil energies and (2) light DM masses which are near or below Cherenkov threshold and thus difficult to probe at Super-K.

Lastly, we stress that the ``effective field theory'' of DD proposed in~\cite{Fitzpatrick:2012ix,DelNobile:2013sia} does not encapsulate the scenario described in this paper and should be extended to include generalized relativistic scattering.


\acknowledgements

IMS is grateful to Yanou Cui and Luca Vecchi for helpful discussions.  The CP$^3$-Origins center is partially funded by the Danish National Research Foundation, grant number DNRF90.  {This work has also been partially funded by the DOE Office of Science and the U.C. Office of the President in conjunction with the LDRD Program at LANL.}

\section*{Appendix}
\label{sec:app}
\subsection{Stopping Effects}
{
The mean free path of the $Y$ particles as they enter the Earth is, $\lambda =\left( n \sigma\right)^{-1}$. The total $Y$-proton cross section is
\be \sigma_{Y-p} \simeq 4\times 10^{-35}~{\rm cm}^{2}~\left(\frac{E_{Y}}{10~{\rm MeV}}\right)^{2} \left(\frac{G_{Y}}{7 \times 10^{4}G_{F}}\right)^{2}.
\ee
%
For a conservative estimate, we take the Earth to be entirely composed of Iron and obtain $n_{N} = \rho_{\oplus}/m_{Fe} \simeq 5.5 \times 10^{22}~{\rm cm}^{-3}$ where $\rho_{\oplus} = 5.5~ {\rm g}~{\rm cm}^{-3}$. Thus the mean free path of $Y$ is
\be \lambda = \frac{1}{n_{N} \sigma_{YN}} \simeq 7000~{\rm km}~\left(\frac{10~{\rm MeV}}{E_{Y}}\right)^{2}~\left(\frac{7 \times 10^{3}G_{F}}{G_{Y}}\right)^{2}
\ee
This can be approximately incorporated into Fig.1 by replacing the flux with $\Phi_{Y} \rightarrow \Phi_{Y} \times \langle \exp \left(-\ell(\theta)/\lambda\right)\rangle$ where the brackets indicate averaging over arrival directions and $\ell(\theta)$ is the chord length an incoming $Y$ particle traverses in the Earth prior to arrival at the detector as a function of the incident angle, $\theta$.
}

\subsection{Higgs portal Example}
\label{appB}
In a Higgs portal implementation, we have $\mathscr{L} \supset b |\phi|^{2} |H|^{2} + g \phi \overline{Y} Y$. Under the assumption that the scalar potential breaks the global $U(1)$, it will develop a vacuum expectation value and mass mix with the SM Higgs, $H$. The precise value of the mixing angle depends on the scalar sector, which is a model-dependent feature. As is well-known, the Higgs portal is constrained by the invisible branching ratio of the Higgs, which is constrained to be $\lesssim 26\%$~\cite{Giardino:2013bma} (though see~\cite{Zhou:2014dba} for weaker, direct constraints). In our case, the Higgs has two new contributions to its invisible width, $h \rightarrow \phi \phi$ and $h \rightarrow \overline{Y}Y$, with widths
\bea \Gamma(h \rightarrow \phi \phi) &=& \frac{b^{2}v_{EW}^{2}}{8 \pi m_{h}}\left( 1- \frac{4 m_{\phi}^{2}}{m_{h}^{2}}\right)^{1/2}, \\
\Gamma(h \rightarrow \overline{Y} Y) &=& \frac{g^{2} \sin^{2} \theta}{8 \pi m_{h}}\left( 1- \frac{4 m_{Y}^{2}}{m_{h}^{2}}\right)^{1/2}
\eea
 We find that these two processes respectively require, $b \lesssim 7.1 \times 10^{-3}$ and $g \sin\theta \lesssim 1.7 \times 10^{-2}$. 

 For sufficiently large $\phi$ mass, we can parameterize the $Y$-$N$ interaction as contact and write the vertex as $G_{Y} (\overline{N} N)( \overline{Y}Y)$ where
\be G_{Y} = \frac{(g \sin \theta) f_{N}}{ m_{\phi}^{2}} \simeq 7 \times 10^{3}~G_{F}~\left( \frac{g \sin \theta}{10^{-2}}\right) \left(\frac{0.2~{\rm GeV}}{m_{\phi}}\right)^{2}
\ee
where $f_{N} =0.345$~\cite{Cline:2013gha}  is the effective coupling of the $\phi$ particles to nucleons. We conclude that the Higgs portal is a viable possibility for the interactions of $Y$ and nuclei. Note further that the coupling to leptons is naturally Yukawa suppressed.

\subsection{Gauged Baryon Number Example}
\label{appC}

Interactions of the type described by $\mathscr{L}_{V}$ arise in models of gauged baryon number~\cite{Pospelov:2011ha,Pospelov:2012gm,Pospelov:2013rha}, gauged $U(1)_{B-L}$~\cite{Nelson:2007yq,Harnik:2012ni}, and kinetic mixing~\cite{Harnik:2012ni}, though it is only in the first case that couplings to leptons are absent.  A detailed exploration of the allowed parameter space of Eq.~(\ref{eq:model}) is beyond the scope of this work, but see~\cite{Harnik:2012ni} for a useful compendium of constraints.  

In the $U(1)_{B}$ case the dominant coupling to hadrons is explicitly guaranteed by the baryon symmetry. The Lagrangian includes $\mathscr{L}_{B} \supset V^{\mu} \left(g_{q} \overline{q}\gamma_{\mu} q + g_{Y}\overline{Y} \gamma_{\mu} Y\right)$. Light baryonic gauge bosons are constrained by the invisible {decay} width of the $\Upsilon(1 {\rm S})$~\cite{Carone:1994aa,Graesser:2011vj}, which is constrained to have a branching ratio, $\mathcal{BR}\left(\Upsilon(1 {\rm S}) \rightarrow {\rm invisible}\right) < 3\times 10^{-4}$~\cite{Aubert:2009ae}, missing energy from the LHC/Tevatron~\cite{Shoemaker:2011vi} and rare radiative decays of the light mesons~\cite{Tulin:2014tya}. As an example we take $m_{V} < 100$ MeV, where we find the mono-jet constraints from CDF to be strongest. In the limit $m_{Y} \ll m_{V}$, these searches constrain the combination of parameters $g_{q} \sqrt{BR(V_{\mu} \rightarrow \overline{Y}Y)} \lesssim 0.02$.

Mapping again to the effective parameter $G_{Y}$ appearing in Eq.(\ref{eq:diffcx}) we find
\be
G_{Y} = \frac{g_{q}g_{Y}}{m_{V}^{2}}  \simeq 7 \times 10^{3}~G_{F}~\left( \frac{g_{q}g_{Y}}{10^{-4}}\right) \left(\frac{35~{\rm MeV}}{m_{V}}\right)^{2},
\ee
and thereby conclude the ``baryonic portal'' is another viable model yielding the phenomenology outlined in the main text of the paper.

\bibliographystyle{JHEP}

\bibliography{nu}

\end{document}